\documentclass{aastex}          
\usepackage{spr-astr-addons}    

\begin{document}
%
\title{Unexplained Spectral Phenomena in the Interstellar Medium: an introduction}

\shorttitle{Unexplained spectral phenomena}
\shortauthors{<Kwok>}

\author{Sun Kwok\altaffilmark{1}}\email{skwok@eoas.ubc.ca} 

\altaffiltext{1}{Department of Earth, Ocean, and Atmospheric Sciences, University of British Columbia, Vancouver, Canada; corresponding author:skwok@eoas.ubc.ca}

\noindent{Astrophysics and Space Science (in press)\\

\begin{abstract}

There exists a number of astronomical spectral phenomena that have remained unidentified after decades of extensive observations.   The diffuse interstellar bands, the 220 nm feature, unidentified infrared emission bands, extended red emissions, and 21 and 30 $\mu$m emission features are seen in a wide variety of astrophysical environments.  The strengths of these features suggest that they originate from chemical compounds made of common elements, possibly organic in nature.  
The quest to understand how such organic materials are synthesized and distributed across the Galaxy represents a major challenge to our understanding of the chemical content of the Universe.

\end{abstract}


\section{Historical background}

In 1814, Joseph Fraunhofer observed and tabulated 473  dark lines in the spectrum of the Sun.  By comparing the Fraunhofer lines with the bright color lines observed in heated chemical elements, Gustav Kirchhoff and Robert Bunsen identified the solar lines as originating from elements sodium, calcium, magnesium, iron, chromium, nickel, barium, copper, and zinc.  This led to the realization that the Sun is made of the same chemical elements as those of the Earth and marked the demise of  the Aristotelian  concept of celestial objects being composed of ether.

In 1868, Norman Lockyer discovered a bright yellow-orange line in the solar spectrum, which he suggested to be due to a new element and named it ``helium''.  The existence of this new element was confirmed by the discovery of its terrestrial counterpart in 1895.
In 1869, a green line was found in the spectrum of the corona of the Sun during the total solar eclipse, and this line was also thought to be a new element and was named ``coronium''.  
In 1864, William Huggins found a bright green emission line in the spectrum of the planetary nebula NGC 6543.  Since this line did not match any lines from known chemical elements, it was suggested to be due to a new element ``nebulium''.  It was not until the early 20th century that ``coronium'' was identified as electronic transitions from ionized iron (Fe$^{13+}$) due to the high temperature in the solar corona, and ``nebulium'' was identified as forbidden lines of ionized oxygen (O$^{++}$) arising under low-density nebular conditions.  These applications  of quantum theory of atoms and laboratory atomic spectroscopy to explain astronomical observations represented the beginning of the modern discipline of astrophysics.  

In the early 21st century, we are facing similar challenges in the form of a number of unexplained spectral phenomena in the interstellar medium.  The diffuse interstellar bands (DIBs) were discovered in 1922, when two optical absorption lines of interstellar origin were seen in the spectra of stars  \citep{heger1922}.  As of 2021, over 500 DIBs in the  ultraviolet, visible, and infrared wavelength regions have been cataloged along the line of sights of over 100 stars.  

The 220 nm ultraviolet feature was discovered in 1965 \citep{stecher1965}.  This feature is seen in the extinction curves of many stars, with characteristically consistent profiles and peak wavelengths.  Its wide presence suggests that the carrier is a common constituent of the diffuse interstellar medium.

The extended red emission (ERE) is a broad ($\Delta\lambda \sim$ 80 nm) emission band peaking between 650 and 800 nm, first discovered in the reflection nebula HD44179 \citep{cohen1975}.  ERE has  been detected in reflection nebulae, dark nebulae, cirrus clouds, planetary nebulae, H~{\sc ii} regions, diffuse interstellar medium, and haloes of galaxies.  It is commonly attributed to photoluminescence powered by far UV photons. It is estimated that $\sim$4\% of the energy absorbed by interstellar dust at $\lambda<$0.55 $\mu$m is emitted in the form of the ERE.

A family of unidentified infrared emission (UIE) features at 3.3, 6.2, 7.7, 8.6, and 11.3 $\mu$m was discovered in the spectrum of the planetary nebula NGC 7027 \citep{russell1977}.
The  3.3 $\mu$m feature was first identified as the C$-$H stretching mode of aromatic compounds by  \citet{knacke1977}.  
The organic origin of the UIE bands was extensively discussed by \citet{duley1981}, who assigned the 3.3 and 11.3 $\mu$m features to graphitic (aromatic) materials.  

Also present in astronomical spectra are emission features around 3.4 $\mu$m, which arise from symmetric and anti-symmetric C$-$H stretching modes of methyl and methylene groups \citep{deMuizon1990}.  The bending modes of these groups also manifest themselves at 6.9 and 7.3 $\mu$m. 
In addition, there are unidentified emission features at 15.8, 16.4, 17.4, 17.8, and 18.9 $\mu$m.
The emission bands themselves are often accompanied by strong, broad emission plateaus features at 6$-$9, 10$-$15, and 15$-$20 $\mu$m.  The first two plateau features have been identified as superpositions of in-plane and out-of-plane bending modes emitted by a mixture of aliphatic side groups attached to aromatic rings \citep{kwok2001}. 
This collection of features in the UIE family has been observed in planetary nebulae, reflection nebulae, novae, H{\sc ii} regions, and galaxies.

The unidentified infrared emission feature around 30 $\mu$m was discovered from {\it Kuiper Airborne Observatory} observations \citep{forrest1981}. It was first seen in carbon-rich asymptotic giant branch stars, planetary nebulae, and proto-planetary nebulae.
Among planetary nebulae in the Magellanic Clouds, about half of them possess the 30 $\mu$m feature \citep{bernard-salas2009}.

The 21 $\mu$m emission feature was first discovered in {\it Infrared Astronomical Satellite} Low Resolution Spectroscopic survey  \citep{kwok1989}.  The feature peaks around 20.1 $\mu$m and shows a broad ($\sim$ 2 $\mu$m) and smooth profile.  The 21 $\mu$m feature is almost always accompanied by the  30 $\mu$m feature.  The 21 $\mu$m feature is primarily observed in carbon-rich post-asymptotic-giant-branch stars.  From the spectral energy distribution of 21 $\mu$m sources, it is found that the 21 and 30 $\mu$m features can carry respectively up to 8 and 20\% of the total energy output of the objects \citep{hrivnak2000}.  

\section{Distribution in the Universe}

Although these unexplained spectral phenomena were first discovered in the spectra of stars, they are not isolated phenomena as they are observed in a wide range of celestial objects throughout the Universe.
DIBs have been detected in external galaxies with redshifts up to 0.5 \citep{sarre2006}.  The 220 nm feature has been detected in interplanetary dust particles in the Solar System \citep{bradley2005} as well as in distant galaxies with redshift $>$2 \citep{elias2009}.
A survey of 150 galaxies by the {\it AKARI} satellite found that  $\sim$0.1\% of the total energy of the parent galaxies is emitted through the 3.3 $\mu$m UIE band \citep{imanishi2010}. In some active galaxies, up to 20\% of the total luminosity of the galaxy is emitted in the UIE bands \citep{smith2007}.  
The 3.4 $\mu$m aliphatic feature has been detected in absorption in ultraluminous infrared galaxies  \citep{mason2004, risaliti2006}.  
The detection of UIE bands in high-redshift galaxies  \citep{teplitz2007} and quasars \citep{lutz2007} implies that organic compounds were widely present as early as 10 billion years ago.  This suggests that abiological synthesis of complex organics was active through most of the history of the Universe.

\section{Chemical nature of the carriers}\label{carriers}

Because of the strengths and ubiquitous nature of the features, the carrier must be made of common, abundant elements, with the element carbon probably  playing a major role.  While the DIBs are commonly believed to be due to electronic transitions of gas-phase carbon-based molecules, the carrier of the 220 nm feature is more likely to be a carbonaceous solid such as amorphous carbon \citep{Mennella1998}, carbon onions (Iglesias-Groth 2004), hydrogenated fullerences (Cataldo \& Iglesias-Groth 2009), or polycrystalline graphite (Papoular \& Papoular 2009).  
Among the hundreds of DIBs, only two  (963.2 and 957.7 nm) have been positively identified as originating from ionized fullerene \citep[C$_{60}^+$,][]{foing1994, campbell2015}.  Two weaker lines of C$_{60}^+$ at 942.8 and 936.6 nm have also been suggested to have counterparts in DIBs \citep{walker2015}.

A variety of chemical structures have been suggested as the carriers of the UIE bands.  These include polycyclic aromatic hydrocarbon (PAH) molecules \citep{leger1984, alla1989}, small carbonaceous molecules \citep{bernstein2009}, hydrogenated amorphous carbon (HAC), soot and carbon nanoparticles \citep{hu2008}, quenched carbonaceous composite particles \citep[QCC,][]{sakata1987}, coal and kerogen  \citep{papoular1989, papoular2001}, petroleum fractions \citep{cataldo2002}, and mixed aromatic/aliphatic organic nanoparticles  \citep[MAON,][]{kz2013}.  

Proposed carriers for the 21 $\mu$m feature include hydrogenated fullerenes (Webster 1995), titanium carbide \citep{vonhelden2000}, silicon carbide \citep{speck2004}, and thiourea groups attached to aromatic/aliphatic structures \citep{pap11}.  
Since ERE is the result of photoluminescience, the carrier is likely a semiconductor with a nonzero band gap.
Other proposed carbon-based carriers include QCC \citep{sakata1992}, C$_{60}$ \citep{webster1993}, and nanodiamonds \citep{chang2006}.

\section{Future outlook}

In this topical collection, the recent developments in our understanding of the ERE and the 21/30 $\mu$m features are reviewed by  \citet{witt2020}, and \citet{volk2020} respectively.  The possible ERE-DIB connection is also discussed by \citet{witt2020}.  
The scenario of  coal and petroleum related compounds as carriers of UIE is discussed by  \citet{cataldo2020}.   The possibilities of hydrogenated fullerenes as carriers of DIB, ERE, and 220 nm features are reviewed by \citet{zhang2020}.   

Is it possible that these unexplained spectral phenomena be related to each other?  The identification of two DIBs with C$_{60}^+$ suggests the possibility that the DIB carrier molecules may be breakdown products of large, complex organic compounds, such as MAONs or other amorphous hydrocarbons.  The presence of the 8 and 12 $\mu$m UIE plateau features in fullerene sources also suggests that fullerenes and UIE carriers may share common precursors.

While we do not know when and where the carriers of DIBs and 220 nm features are synthesized, the observations of UIE bands, ERE, and 21/30 $\mu$m features in the circumstellar environment dictate that the chemical synthesis time scale is constrained by the evolutionary and dynamical time scale of the circumstellar envelopes.   For example, UIE bands are observed to emerge over time scales of $\sim$10$^3$ years in the proto-planetary nebulae phase \citep{kwok1999}, and over time scales of weeks in novae \citep{helton2011}.  This suggests that the synthesis of the UIE carriers is extremely efficient.  How such synthesis can occur so rapidly under low-density conditions is not understood by our current chemical models.

Could the carriers of these unexplained spectral phenomena be new chemical compounds unobserved on Earth, as in the case of the discovery of helium in the Sun, or could they be the result of unusual physical environment as in the case of ``coronium'' and ``nebulium''?
Further development in laboratory spectroscopy will hold the key to the identification of the carriers. 

\citet{snow2014} has suggested that  DIBs alone represent the largest reservoir of organic material in the Galaxy, so no matter what the exact nature of the carriers is, they must be an important constituent of the Cosmos.   The resolution of these spectral mysteries may bring about a new view of our understanding of the chemical content of the Universe.

\acknowledgments

This work is supported by a grant from the Natural and Engineering Research Council of Canada.


\begin{thebibliography}{}

\bibitem[Allamandola, Tielens, and Barker(1989)]{alla1989}
Allamandola,  L.J., Tielens, A.G.G.M., Barker, J.R.:  Astrophys. J. Suppl. {\bf 71}. 733 (1989)


\bibitem[Bernard-Salas et al.(2009)]{bernard-salas2009} 
Bernard-Salas, J., Peeters, E., Sloan, G. C., et al. , \apj, {\bf 699}. 1541 (2009)

\bibitem[Bernstein and Lynch(2009)]{bernstein2009}
Bernstein, L.S., Lynch, D.K.:  Astrophys J {\bf 704}. 226 (2009) 

\bibitem[Bradley et al.(2005)]{bradley2005}
Bradley, J., Dai, Z.R., Erni, R., Browning, N., Graham, G., Weber, P., Smith, J., Hutcheon, I., Ishii, H., Bajt, S., Floss, C., Stadermann, F., Sandford, S.: Science   {\bf 307}. 244 (2005)

\bibitem[Campbell et al.(2015)]{campbell2015}
Campbell, E.K., Holz, M., Gerlich, D., Maier, J.P.:  Nature {\bf 523}. 322  (2015)

\bibitem[Cataldo, Feheyan, and Heymann(2002)]{cataldo2002}
Cataldo, F., Keheyan, Y., Heymann, D.:   International J. of Astrobiology {\bf 1}. 79 (2002)

\bibitem[Cataldo, Garc\'{i}a-Hern\'{a}ndez and Manchado(2020)]{cataldo2020}
Cataldo, F., Garc\'{i}a-Hern\'{a}ndez, D. A., \& Manchado, A.: \apss  {\bf 365}. 81 (2020)

\bibitem[Chang, Chen, \& Kwok(2006)]{chang2006}
Chang, H.-C., Chen, K., Kwok, S.:  Astrophys. J. Lett. {\bf 639}. L63 (2006)



\bibitem[Cohen et al.(1975)]{cohen1975} 
Cohen, M., Anderson, C. M., Cowley, A., et al.: \apj\ {\bf 196}. 179 (1975)

\bibitem[Duley \& Williams(1981)]{duley1981}
Duley, W.W., Williams, D.A.: Mon. Not. R. Astron. Soc. {\bf 196}. 269 (1981)


\bibitem[El\'{i}asd\'{o}ttir et al.(2009)]{elias2009}
El\'{i}asd\'{o}ttir, \'{A}. et al.: Astrophys. J. {\bf 697}. 1725 (2009)

\bibitem[Foing \& Ehrenfreund(1994)]{foing1994}
Foing, B.H., Ehrenfreund, P.: Nature {\bf 369}. 296 (1994)

\bibitem[Forrest, Houck, and McCarthy(1981)]{forrest1981} 
Forrest, W. J., Houck, J. R., \& McCarthy, J. F. 1981, \apj\ {\bf 248}. 195 (1981)

\bibitem[Heger(1922)]{heger1922}
Heger, M. L.: Lick Observatory Bulletin {\bf 10}. 141 (1922)



\bibitem[Helton et al.(2011)]{helton2011}
Helton, L. A., Evans, A., Woodward, C. E.,  Gehrz, R. D.: in Proceedings of the EAS Publications Series {\bf 46}.  407

\bibitem[Hrivnak, Volk, and Kwok(2000)]{hrivnak2000} 
Hrivnak, B. J., Volk, K., Kwok, S.: \apj\ {\bf 535}. 275 (2000)

\bibitem[Hu and Duley(2008)]{hu2008}
Hu, A., Duley, W.W.: Astrophys. J. Lett. {\bf 677}.  L153 (2008)

\bibitem[Imanishi et al.(2010)]{imanishi2010}
Imanishi, M., Nakagawa, T., Shirahata, M., Ohyama, Y., Onaka, T.:  Astrophys. J. {\bf 721} 1233 (2010)

\bibitem[Jourdain de Muizon, D'Hendecourt and Geballe(1990)]{deMuizon1990} 
Jourdain de Muizon, M., D'Hendecourt, L. B.,  Geballe, T. R.:  \aap\ {\bf235}. 367 (1990)

\bibitem[Knacke(1977)]{knacke1977}
Knacke, R.F.:  Nature {\bf 269}. 132 (1977)

\bibitem[Kwok \& Zhang(2013)]{kz2013}
Kwok, S., Zhang, Y.: Astrophys. J. {\bf 771}.  5 (2013)

\bibitem[Kwok, Volk, and Bernath(2001)]{kwok2001}
Kwok, S., Volk, K., Bernath, P.: Astrophys. J. Lett. {\bf 554}.  L87 (2001)


\bibitem[Kwok, Volk, and Hrivnak(1989)]{kwok1989} 
Kwok, S., Volk, K. M., \& Hrivnak, B. J.: Astrophys. J. Lett. {\bf 345}. L51 (1989)





\bibitem[Kwok, Volk, \& Hrivnak(1999)]{kwok1999} 
Kwok, S., Volk, K., \& Hrivnak, B. J.: \aap\ {\bf 350}. L35 (1999)

\bibitem[L\'{e}ger and Puget(1984)]{leger1984}
L\'{e}ger, A., Puget, J.L.: Astron. Astrophys. {\bf 137}. L5 (1984)

\bibitem[Lutz et al.(2007)]{lutz2007}
Lutz, D., et al.:
Astrophys. J. Lett. {\bf 661}.  L25 (2007)

\bibitem[Mason et al.(2004)]{mason2004}
Mason, R.E., Wright, G., Pendleton, Y., Adamson, A.:
Astrophys. J. {\bf 613}.  770 (2004)

\bibitem[Mennella et al.(1998)]{Mennella1998}
Mennella, V., Colangeli, L., Bussoletti, E., Palumbo, P., Rotundi, A.: Astrophys. J. Lett. {\bf 507}. L177 (1998)



\bibitem[Papoular(2001)]{papoular2001}
Papoular, R.: Astron. Astrophys. {\bf 378}. 597 (2001)

\bibitem[Papoular(2011)]{pap11}
Papoular, R.: Mon. Not. R. Astron. Soc. {\bf 415}. 494 (2011)

\bibitem[Papoular \& Papoular(2009)]{pap09}
Papoular, R.J., Papoular, R.: Mon. Not. R. Astron. Soc. {\bf 394}. 2175 (2009)

\bibitem[Papoular et al.(1989)]{papoular1989}
Papoular, R., Conrad, J., Giuliano, M., Kister, J., Mille, G.: 
Astron. Astrophys. {\bf 217}.  204 (1989)




\bibitem[Risaliti et al.(2006)]{risaliti2006}
Risaliti, G., Maiolino, R., Marconi, A., et al. \mnras\ {\bf 365}. 303 (2006)

\bibitem[Russell, Soifer, and Willner(1977)]{russell1977}
Russell, R.W., Soifer, B.T., Willner, S.P.:
Astrophys. J. Lett. {\bf 217} L149 (1977)

\bibitem[Sakata et al.(1987)]{sakata1987}
Sakata, A., Wada, S., Onaka, T., Tokunaga, A.T.:
Astrophys. J. Lett. {\bf 320}. L63 (1987)

\bibitem[Sakata etal(1992)]{sakata1992}
Sakata, A. etal: Astrophys. J. {\bf 393}. L83 (1992)

\bibitem[Sarre(2006)]{sarre2006}
Sarre, P.J.:  J. Molecular Spectroscopy {\bf 238(1)}. 1 (2006)

\bibitem[Smith et al.(2007)]{smith2007}
Smith, J.D.T., et al.: 
Astrophys. J. {\bf 656}.  770 (2007)

\bibitem[Snow(2014)]{snow2014}
Snow, T.P.: in IAU Symposium 297: the Interstellar Diffuse Bands, eds. J. Cami \& N.L.J. Cox, CUP, p. 3 (2014)

\bibitem[Speck and Hofmeister(2004)]{speck2004} 
Speck, A. K., Hofmeister, A. M.:  \apj\ {\bf 600}. 986 (2004)

\bibitem[Stecher(1965)]{stecher1965} 
Stecher, T. P.: Astrophys. J.  {\bf 142}. 1683 (1965)

\bibitem[Teplitz et al.(2007)]{teplitz2007}
Teplitz, H.I., et al.: 
Astrophys. J. {\bf 659}. 941 (2007)

\bibitem[Walker et al.(2015)]{walker2015}
Walker, G. A. H., Bohlender, D. A., Maier, J. P., Campbell, E. K.: Astrophys. J. Lett. {\bf 812}. L8 (2015)


%

\bibitem[Volk, Sloan, and Kraemer(2020)]{volk2020}
Volk, K., Sloan, G. C., Kraemer, K. E.: \apss\ {\bf 365}. 88 (2020)

\bibitem[von Helden et al.(2000)]{vonhelden2000}
von Helden, G. et al.:
Science {\bf 288}. 313 (2000)

\bibitem[Webster(1993)]{webster1993}
Webster, A.: Mon. Not. R. Astron. Soc. {\bf 264}. L1 (1993)

\bibitem[Webster(1995)]{webster1995}
Webster, A.: 
Mon. Not. R. Astron. Soc.  {\bf 277}. 1555 (1995)


\bibitem[Witt and Lai(2020)]{witt2020}
Witt, A. N., Lai, T. S.-Y.:  \apss\ {\bf 365}. 58 (2020)

\bibitem[Zhang, Sadjadi, and Hsia(2020)]{zhang2020}
Zhang, Y., Sadjadi, S., Hsia, C.-H.: \apss\ {\bf 365}. 67 (2020)

\end{thebibliography}
\end{document}